\documentclass[twocolumn, prl]{revtex4}
\usepackage{graphicx}

\begin{document}
\title{Harmonic inversion helps to beat time-energy uncertainty relations.}
\author{Zbyszek P. Karkuszewski}

\affiliation{
Institute of Physics, Jagiellonian University, ul. Reymonta 4, 30-059 Cracow, 
Poland
}

\date{\today}

\begin{abstract}
It is impossible to obtain accurate frequencies from time signals of a very
short duration. This is a common believe among contemporary physicists.
Here I present a practical way of extracting energies to a high precision from
very short time signals produced by a quantum system. The product of time span 
of the signal and the precision of found energies is well bellow the limit 
imposed by the time-energy uncertainty relation.
\end{abstract}

\maketitle

\section{Introduction}
Many methods of obtaining frequencies or energies from time signals have limited
resolution. This limitation is often expressed in the form of a time-energy 
uncertainty relation. For instance, the width $\Delta E$ of spectral lines
in atomic spectroscopy is determined by life time $\tau$ of the excited 
atoms
$$
\Delta E \tau = \hbar.
$$
Another example is a Fourier transform applied to a discrete time signal in
the interval of length $T$. The grid step in a frequency domain and thus the 
resolution of the method is given by $\Delta \omega = 2\pi/T$, which in turn 
yields
$$
\Delta \omega T = 2\pi.
$$
All relations of this kind may give a wrong impression that something more
fundamental is the source of the uncertainties -- a principle that one
cannot measure energies to an arbitrary precision in a very short time.
Such an intuition is common if the system subjected to energy measurement is 
quantum. 

Uncertainty principles in quantum mechanics are a mathematical 
consequence of the following theorem.
Two self-adjoint operators $\hat A$ and $\hat B$ defined on the same 
Hilbert space necessarily obey the relation
\begin{equation}
\Delta A\Delta B \ge \frac{1}{2}|\langle[\hat A,\hat B]\rangle|,
\label{qur}
\end{equation}
where $[...,...]$ is a commutator, $\langle ...\rangle$ stands for quantum 
average in a given state from the domain of $\hat A$ and $\hat B$, 
and $(\Delta Z)^2 \equiv \langle \hat Z^2\rangle-\langle \hat Z \rangle ^2$.
In the case of position $\hat x$ and momentum $\hat p$ operators (\ref{qur}) 
gives
\begin{equation}
	\Delta x \Delta p \ge \frac{\hbar}{2}.
\label{xpr}
\end{equation}
Notice that $\Delta x$ and $\Delta p$ are completely determined by the state
of the system being measured and have nothing to do with an accuracy of 
the measuring apparatus. (\ref{xpr}) holds even if this accuracy is infinite.
The correct approach to the quantum time-energy uncertainties can be found 
in~\cite{Asher}.
The theorem (\ref{qur}) does not apply to the case of time and energy,
simply because time is not an operator, it is just a parameter in quantum
mechanics. One gains nothing forcing the idea that a parameter is a special 
kind operator because (\ref{qur}) gives zeros on both sides and does not
provide grounds for existence of any time-energy uncertainty relation. 
Indeed, it has been shown how to precisely measure energy in an
arbitrarily short time \cite{Aharonov1}. Some methods of the energy measurement 
have their own limitations \cite{Aharonov2} reflected in loss of accuracy 
when applied for very short time signals. Here I use the method free of such
inconveniences. 

Let us setup some general issues before getting to the details of the
method. 
Suppose that a continuous signal $c(t)$ is given in a finite time interval 
\begin{equation}
c(t) = \sum_{k=1}^K d_k e^{-i\omega_k t}, \qquad \mbox{for}\quad 0\le t \le T,
\label{sigc}
\end{equation}
where $K$ is finite, $d_k$ is a real amplitude and $\omega_k$ is a real 
frequency. Complex frequencies will be considered in future. 
The task is to find unknown amplitudes and frequencies of 
$c(t)$. What one can see is that $c(t)$ is an analytic function of
time $t$ and, as such, can be uniquely extended beyond the interval $(0,T)$.
This means that, in principle, even for tiny $T$ it is possible to get all
amplitudes and frequencies to a high precision from (\ref{sigc}). 
Unfortunately it seems difficult to solve this nonlinear problem analytically
and numerical methods cannot handle continuous signals due to infinite number
of data points.  One way of getting around this problem is to take only finite 
number of points at cost of loss of uniqueness of the extension.
However, as will be shown later, this drawback is usually not severe for 
practical purposes. From now on I shall assume that the signal $c(t)$ is
known only at $N+1$ equidistant time points $t_n=n\delta t$ for $n=0,...,N$
and $t_N=T$. 
(\ref{sigc}) can be rewritten as a set of $N+1$ equations
\begin{equation}
\sum_{k=1}^K d_k e^{-i\omega_k t_n} = c_n,
\label{sigd}
\end{equation}
where $c_n\equiv c(n\delta t)$. This set has $2K$ real unknowns 
($K$ amplitudes and $K$ frequencies) so the number of equations $N+1$ has to
be equal or greater than $K$. This condition would have opened the possibility
of existence of a unique solution if the equations were linear in $\omega_k$ 
and $d_k$. It is not the case here. There will always be infinite number of solutions
to (\ref{sigd}) if the set is self-consistent and no solutions otherwise.
For reasons explained in the next sections it will be required that $N\ge K$.
Notice that discrete Fourier transform method assumes the grid of $K$ 
equidistant frequencies and solves (\ref{sigd}) only for $d_k$ as a linear 
set of
equations. The more challenging task of solving (\ref{sigd}) for
amplitudes and frequencies is performed by, so called, harmonic inversion 
method.

\section{Intervals of unique solutions}
The two functions $e^{i\omega n \delta t}$ and $e^{i\omega' n \delta t}$ have 
the same values at equidistant time points $n \delta t$, $n=0,...,N$ if
$$
\omega' = \omega+ l\frac{2\pi}{n\delta t}.
$$
The integer number $l$ has to be chosen in such a way that the fraction
$l/n$ is integer as well. This implies that $l=mN!$, where $m$ is integer
and 
\begin{equation}
\omega' = \omega + m\frac{2\pi N!}{\delta t}.
\label{uniq}
\end{equation}
The time discretization of the signal $c(t)$ , the transition from 
(\ref{sigc}) to (\ref{sigd}), happens at cost of the loss of uniqueness of 
solutions. However, as can be seen from (\ref{uniq}) solutions are unique 
in finite intervals in frequency domain. Frequency $\omega$ is unique
in the interval 
$$
\left( \omega -\frac{2\pi N N!}{T}, \omega + \frac{2\pi N N!}{T}\right),
$$
which can be made large by increasing $N$ and/or decreasing $T$.
Conversely, if one posses prior knowledge about the range of frequencies
 in a signal it is straightforward to design a sampling step $\delta t$ to
 extract all frequencies in a unique way.

\section{Harmonic inversion}
The key idea behind the harmonic inversion method is to replace the original 
nonlinear problem (\ref{sigd}) with an eigenvalue problem of an operator, as 
in (\ref{gep}). The presentation of the method in this section follows that 
in \cite{Taylor2}.

Lets start with a normalized quantum state $|\Phi_0\rangle$.  
The evolution of this state is generated by unitary evolution operator 
$\hat U(\delta t)$ and $|\Phi_n\rangle = \hat U^n(\delta t)|\Phi_0\rangle$. 
The states $|\Phi_n\rangle$ are so important that they were given a name of 
Krylov states. For every time signal in (\ref{sigd}) there exists such an 
evolution by time $\delta t$ operator $\hat U$ that the signal can be
viewed as an autocorrelation function
\begin{equation}
c_n = \langle \Phi_0| \Phi_n\rangle.
\label{auto}
\end{equation}
Krylov states in (\ref{auto}) can be written in the orthonormal basis of 
eigenvectors of $\hat U$ defined by $\hat U|u_k\rangle = u_k|u_k\rangle$. 
Namely $|\Phi_n\rangle = \sum_{k=1}^K \alpha_k u_k^n|u_k\rangle$, where 
$\alpha_k$ stands for a time independent complex number. (\ref{auto}) in the 
new form appears as
\begin{equation}
c_n = \sum_{k=1}^K |\alpha_k|^2 u_k^n.
\label{nauto}
\end{equation}
Comparing (\ref{nauto}) and (\ref{gep}) one can see that eigenvalues 
$u_k = e^{-i\omega_k \delta t}$ and $|\alpha_k|^2 = d_k$. It is enough 
to find eigenvalues of $\hat U$ in order to obtain all frequencies $\omega_k$ 
in the signal $c_n$. These frequencies multiplied by $\hbar$ can be viewed as 
eigenenergies of a hypothetical Hamiltonian governing the evolution of a
system in the initial state $|\Phi_0\rangle$. Of course, one can start
from the autocorrelation function (\ref{auto}) and find the eigenenergies
of the real system. 

It turns out that the matrix elements of $\hat U$ in the basis of Krylov states 
can be expressed in terms of $c_n$ alone, i.e. without explicit reference to 
the states $|\Phi_n\rangle$
\begin{equation}
U_{ij}\equiv \langle \Phi_i | U| \Phi_j\rangle = 
\langle \Phi_0| U^{j-i+1} | \Phi_0\rangle = c_{j-i+1} 
\label{me}
\end{equation}
where $i,j=0,..., N-1$. The negative 
indices of $c$ in the equation above introduce no complication since 
$c_{-n}=c^*_n$. To obtain $K$ eigenvalues the dimension $N$ of the matrix 
$U$ must be equal or greater than $K$. It means that the number of
complex signal points $c_n$ required by the method exceeds the half of the 
number of unknowns. The Krylov vectors $|\Phi_n\rangle$ are normalized but not 
orthogonal.
Thus the eigenequation for matrix $U$ in the Krylov representation
(\ref{me}) takes the form
\begin{equation}
U |u_k\rangle = u_k S |u_k\rangle,
\label{gep}
\end{equation}
where $S$ is a matrix of scalar products $S_{ij}\equiv\langle
\Phi_i|\Phi_j\rangle=c_{j-i}$ with $i,j=0,...,N-1$. 

The harmonic inversion method consists of two stages. First is to solve
generalized eigenvalue problem (\ref{gep}), where all matrix elements
are expressed in terms of $c_n$, in order to get all frequencies. 
Second, when all frequencies in (\ref{sigd}) are known, it is enough to
solve linear set of equations for the amplitudes $d_k$. 
There are some practical difficulties in proceeding with the first stage. 
For instance, numerical algorithms fail in finding the eigenvalues $u_k$ 
if the signal duration $T$ is small and the number of frequencies $K>4$. 
The second stage is straightforward and will be skipped in this work. It
will be assumed that all real amplitudes $d_k$ are equal and the signal
is normalized to unity, i.e. $c_0 = 1$.

In the next section we will see that the crucial role in the numerical
approach to (\ref{gep}) is played by the smallest positive eigenvalue of $S$.

\section{Properties of matrix S}
\begin{figure}[htb]
\includegraphics*[width=8.6cm]{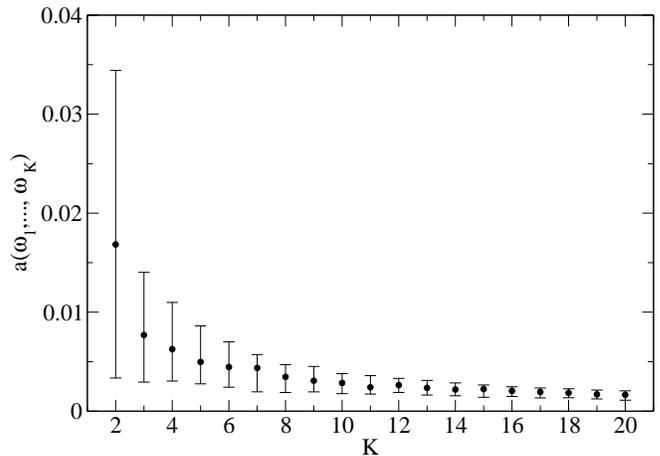}
\caption{Statistical properties of $a(\omega_1,...,\omega_K)$. For every $K$ 
100 sequences of $K$ frequencies were randomly generated. Each frequency has
been uniformly drawn from the interval $(0.5,1.0)$. For each
sequence $a(\omega_1,...,\omega_K)$ was computed. Circles denote the most
probable value of $a(\omega_1,...,\omega_K)$ and vertical bars embrace 90\%
of sequences.} 
\label{Fig1}
\end{figure}

One way of dealing with the generalized eigenvalue problem is to 
reduce it to the ordinary eigenvalue problem by multiplying both sides of 
(\ref{gep}) by the inverse of the Hermitian Toeplitz matrix $S$. However, 
matrix $S$ is singular unless $N=K$ and the inverse does not exist.
Indeed, recall that every vector $|\Phi_n\rangle$ can be decomposed into a 
superposition of $K$ linearly independent eigenvectors of $\hat U$. 
Therefore the rank of the matrix $S$ is $K$. Additionally, those nonzero
eigenvalues are positive because a scalar product 
$\langle \varphi|\varphi \rangle > 0$ for any nonzero state 
$|\varphi \rangle$ in a Hilbert space.
Suppose that columns of a $N\times K$ matrix $G$ are represented by 
eigenvectors of matrix $S$ to the positive eigenvalues. In order to 
restrict (\ref{gep}) to the range of $S$ we define $U'\equiv G^\dagger U G$, 
$S'\equiv G^\dagger S G$ and $|u_k'\rangle \equiv G^{-1}|u_k\rangle$.
$S'$ is diagonal and positive defined thus the ordinary eigenvalue problem
converts to
\begin{equation}
S'^{-1} U' |u_k'\rangle  = u_k |u_k'\rangle .
\label{ep}
\end{equation}
Construction of matrix $G$ is possible only if one can extract all positive
eigenvalues of $S$. This task becomes hopeless if the smallest eigenvalue
of $S$ cannot be distinguished from zero due to limited accuracy.
The exact analytical expression for the smallest eigenvalue of a Toeplitz 
matrices is not known yet.
Here is the approximate formula for the magnitude of the smallest 
eigenvalue for the short time span $T$ of the signal  
\begin{equation}
\frac{\lambda_{min}}{KN} \approx [a(\omega_1,...,\omega_K)\Omega T]^{2(K-1)}.
\label{mineig}
\end{equation}
$\Omega$ is the magnitude of the greatest (to the absolute value) 
frequency in the signal. For the derivation of (\ref{mineig}) see the
directions in the Appendix.
The formula is valid only if the shortness
condition is fulfilled, $T\Omega \ll 1$. The expression in square
brackets in (\ref{mineig}) is smaller than 1 and the $\lambda_{min}$ decreases
exponentially fast with increasing $K$. For the estimated magnitude of the 
function $a(\omega_1,...,\omega_K)$ see Fig.\ref{Fig1}.

The eigenvalue $\lambda_{min}$ is additionally 
corrupted by inaccuracies of the signal. The accuracy analysis is presented 
in the next section.

\section{Impact of noise}
From now on it will be assumed that the accurate signal $c_n$ is perturbed
by noise $\eta_n$ and the new signal 
$$
\tilde c_n = c_n + \eta_n.
$$
The noise is limited, $\eta_n \in (-\eta_{max}, \eta_{max})$ for all $n$
and $\eta_{max}\ge 0$.
Matrix $\tilde S$, which is constructed using $\tilde c_n$ instead
of $c_n$ (see the text under (\ref{gep})), remains Hermitian.
It can be shown \cite{golub} that the smallest eigenvalue of $\tilde S$
must obey the inequality
$$
|\tilde \lambda_{min} - \lambda_{min}|\le 2N\eta_{max}+ {\cal O}(\eta_{max}^2).
$$
If $\tilde \lambda_{min}$ is to be distinguished from perturbed zero
eigenvalues the following lower bound on $\lambda_{min}$ arises
\begin{equation}
\lambda_{min}\ge 4N\eta_{max}.
\label{bound}
\end{equation}
This diagnostic condition combined with (\ref{mineig}) provides the limits of 
applicability of the harmonic inversion method for short signals. 

Tracking the inaccuracy propagation when solving (\ref{ep}) one arrives
at the surprising at first sight "certainty relation" on perturbed 
frequencies $\tilde \omega_k$
\begin{equation}
|\tilde \omega_k - \omega_k|T \le \frac{2KN^2}{\lambda_{min}}\eta_{max}.
\label{ina}
\end{equation}
The error concerning frequencies of short signals can be made arbitrarily 
small by reducing the noise amplitude!
Higher order terms in $\eta_{max}$ were dropped in (\ref{ina}).
For time signals of short duration the presence of $\lambda_{min}$ in the
denominator in RHS of (\ref{ina}) is unfavorable. As stated in 
(\ref{mineig}) $\lambda_{min}$ rapidly goes to zero as $K$ increases.
Therefore, to extract many accurate frequencies from a short signal the very
low level of the noise $\eta_{max}$ will be required. 

In numerical experiments the noise is caused by finite precision of floating
point numbers. As an example, the harmonic inversion method was used
on a signal with $K=10$ frequencies drawn from an interval $(0.5,1.0)$,
sampled at $N=14$ points with $T=0.01$ and using 85 digits precision i.e. 
$\eta_{max}=10^{-84}$. The results are shown in Table~\ref{tab:res}.
\begin{table}[htb]
\caption{\label{tab:res}Numerical example. All amplitudes $d_k$ were equal.}
\begin{ruledtabular}
\begin{tabular}{c | c | c}
$\omega_k$ & $\tilde \omega_k$& $|\tilde\omega_k - \omega_k|$\\
\hline
0.50415486481506&0.50415486481507&0.00000000000001\\
0.51315149664879&0.51315149664880&0.00000000000002\\
0.66526505816728&0.66526505819813&0.00000000003085\\
0.71068158128764&0.71068158894354&0.00000000765589\\
0.73253390251193&0.73253391404702&0.00000001153508\\
0.75819833122659&0.75819832230088&0.00000000892572\\
0.79694000270683&0.79694000183578&0.00000000087105\\
0.85043252643663&0.85043252642270&0.00000000001394\\
0.88220469909720&0.88220469909823&0.00000000000103\\
0.93358750757761&0.93358750757763&0.00000000000001\\
\end{tabular}
\end{ruledtabular}
\end{table}
Notice that the Fourier algorithm applied to this signal would give 
the grid step in frequency domain $200\pi$ which is several orders
of magnitude greater than the error in Table~\ref{tab:res}.
In the example above $\lambda_{min}=4.07\times 10^{-78}$.

\section{Disadvantages of the method}
Harmonic inversion method does pretty well when applied to short time signals
provided the level of inaccuracies in the input data is very small.

In principle this level can be kept very low in measurement of quantum
systems. For instance, the measurement of quantum autocorrelation
function squared (probability of finding a system in the initial state at
different times of its evolution) can be arbitrarily precise if done 
simultaneously on many copies of the system. The inaccuracy decreases 
as an inverse square root of the number of the copies. This function involves
$(K-1)K/2$ frequencies rather than $K$ as in (\ref{auto}).

In experimental practice, however, inaccuracies are big and the method might
be useful only for signals with small number of frequencies.


Even if desired accuracy is provided another problem may show up.
The harmonic inversion method as described here involves diagonalization
of $N\times N$ matrix $\tilde S$ and $K\times K$ matrix 
${\tilde S}^{-1}\tilde U$. 
$\tilde S$ matrix is Hermitian Toeplitz, the matrix elements 
are constant along diagonals, and it is enough to store only one row of 
$\tilde S$. Solving a linear set of equations with Toeplitz matrix requires	 
 $\propto N\log^2_2(N)$ operations using superfast algorithms \cite{superfast}.
Perhaps diagonalization of Toeplitz matrices can be also speed up bellow 
$\propto N^3$ operations.
The complexity of the second diagonalization is $\propto K^3$ operations
but here no eigenvectors need to be computed.
Remember that all these operations has to be done with high precision and are,
therefore, relatively slow.

The diagonalization related difficulties has been overcome for long signals 
\cite{Neuhauser,Taylor1,Taylor2}.
For such problems the frequency domain was effectively divided into smaller 
intervals containing fewer frequencies and
the harmonic inversion, known there as filter diagonalization, was applied to 
one interval at the time. Splitting the frequency domain introduces however
its own inaccuracies into the computed frequencies of the signal and filter 
diagonalization methods are said to obey their own time-frequency uncertainty 
relations. The total duration of the signal is limited from bellow by local 
density of frequencies \cite{Taylor2} or by minimal and average frequency 
distance \cite{Neuhauser}. It is not clear yet whether these restrictions
can be invalidated by improved precision of calculations. 

\section{Summary}
In this work I presented the application of the harmonic inversion
method to short time signals. I have demonstrated that the method has no 
fundamental
limitations regarding the length of the signals. It works for arbitrarily short
signals if sufficient accuracy of the input data is provided. 
In particular, it has been shown that it is possible
to extract energies from the short autocorrelation function generated by a 
quantum system. The method can be also applied to short autocorrelation 
functions squared which have a clear experimental interpretation. Its
measurement can be, in principle, carried out in arbitrarily short 
interval of time and still all energies involved in its evolution can be 
extracted to a desired accuracy.

\section{Acknowledgments}
I am grateful to Jacek Dziarmaga, Krzysztof Sacha, Jakub Zakrzewski and
George Zweig for many stimulating discussions.
Work supported by KBN grant 5 P03B 088 21.

\appendix
\section{\label{app}Appendix: Estimate of $\lambda_{min}$}
The direct calculation of $\lambda_{min}$ is hard. What can be done 
without much effort is the prediction of the general form of the expression
for this quantity.
The first step is the observation that since each matrix element of $S$ 
is a sum of $K$ exponents as in (\ref{sigd}) then $S$ can be written as
a sum of $K$ matrices $S= S_1 + S_2 + ... + S_K$, where each matrix $S_k$
depends only on one frequency $\omega_k$. Matrix $S_k$ is Hermitian and has 
only one nonzero eigenvalue $N$ and corresponding eigenvector 
$|s_k\rangle = [1, e^{i\omega_k\delta t}, e^{i\omega_k2\delta t},...
,e^{i\omega_k(N-1)\delta t}]/\sqrt{N}$. Eigenvectors $|s_k\rangle$
form a nonorthogonal basis. Lets construct a $K\times N$ matrix $R$ by
filing its rows with vectors $|s_k\rangle$. Eigenvalues of $RR^\dagger$ 
multiplied by $N$ give the all nonzero eigenvalues of $S$.
The characteristic equation for eigenvalues of $RR^\dagger$
$$
a_K(\lambda/N)^K +...+a_1(\lambda/N) + \mbox{det}(RR^\dagger) = 0
$$
can be expanded at $\lambda = 0$ to the first order to give an
estimate for the smallest eigenvalue
$$
\frac{\lambda_{min}}{NK} \approx -\frac{\mbox{det}(RR^\dagger/K)}{a_1}.
$$
The matrix $RR^\dagger$ above is conveniently divided by $K$ to normalize its
trace to unity.
For $N=K$ matrix $R$ becomes square and
$\mbox{det}(R)\propto (\Omega T)^{K(K-1)/2}$ for $\Omega T \ll 1$. 
Therefore, $\mbox{det}(RR^\dagger/K)\approx (c_1\Omega T)^{K(K-1)}$.
The parameter $a_1$ is given by sum of minors of
$RR^\dagger/K$ and $a_1 \approx (c_2\Omega T)^{(K-1)(K-2)}$. 
$c_1$ and $c_2$ are functions of normalized frequencies $\omega_k/\Omega$. 
Similar approximations are obtained if $N>K$. 
All calculations are summarized by the estimate (\ref{mineig}).

\end{document}